# Towards field-programmable photonic gate arrays

D. Pérez-López*[1,2], A. López-Hernández[2], A. Macho[2], P. Das Mahapatra[1,2], J. Capmany*[1,2]
[1]iPRONICS, Programmable Photonics, Camino de Vera s/, 46022 Valencia, Spain; [2]Universitat Politècnica de València, iTEAM Research Institute, Camino de Vera s/, 46022 Valencia, Spain

**ABSTRACT**

We review some of the basic principles, fundamentals, technologies, architectures and recent advances leading to thefor the implementation of Field Programmable Photonic Field Arrays (FPPGAs).

**Keywords:** programmable photonics, Integrated Optics, Field Programmable Gate Arrays

## 1. INTRODUCTION

Reconfigurable (or programmable) systems are configurations with soft-definable features or, in other words, soft hardware that can be tuned, reshaped, or otherwise altered at will to suit the purposes of their users [1]. Reconfigurability is now commonplace in electronics components, circuits, and systems with the Field Programmable Gate Array (FPGA) device being one of the best-known examples [2]. FPGAs, the paradigm of electronic reconfigurable systems were conceived to compete with Application Specific Integrated circuits (ASICs) and now after almost 30 years since their inception they have practically replaced them in great percentage of applications [3]. The increased complexity that can be achieved fueled by the benefits of Moore's law and cost reductions due to reduced non-recurring engineering costs have propelled FPGA devices to a leading position. This is further sustained by the trend in fusing practical engineering and design principles that can effectively manage the flexibility required in the so-called software-x approaches that include among others software-defined radios [4], software-defined networks [5], cloud computing [6] and data center concepts [7].

Going further one may ask if it makes sense to consider the extension of programmable systems to other application scenarios such as integrated photonics. It is true that this field has not yet reached the degree of integration maturity of electronics in general and FPGAs in particular. Reconfigurable systems are sometimes criticized because they require overhead, which may lower performance and add complexity, possibly reducing reliability. These considerations must be traded-off against their advantages when considering their use for a particular application or field. Nevertheless, they bring a considerable number of benefits, which can be summarized as follows:

1. *Flexible reshaping of finite resources*: A reconfigurable can be considered as a finite collection of resources that can be reshaped at will, leading to advantages in mass extreme customization, reduction of nonrecurring engineering expenses, economic savings through inventory collapse, design rectification and functional update and iterative refinement to accommodate evolution.

2. *Robustness and resilience*: Reconfigurability results in systems that are tolerant to faults and manufacturing defects by exploiting unused configurable resources. Redundancy approaches can be implemented provided that enough spatial resources are available in the programmable system. Furthermore, by software programming it is possible to create self-healing and/or cooperative multitasking systems.

3. *Achieving "x on demand"*: Reconfigurability brings the advantage of creating systems quickly. Prebuilt parts can be personalized rapidly overcoming the long fabrication cycles of dedicated systems or chips. This is especially important in electronic microfabrication.

4. *Infinite resources through timesharing*: Reconfigurable systems can be thought as a set of infinitely re-purposeable components leading to temporal reuse.

Programmable Integrated Photonics (PIP) [8]-[23] is a new paradigm that aims at designing reconfigurable systems on integrated optic chips leveraging from the former generic advantages. In particular, Field Programmable Photonic Gate Arrays (FPPGAs), recently proposed, aim to play a similar role in photonics as FPGAs do in electronics. While high-level operational architecture principles of FPPGAs and FPGAs share common features, basic operation principles are substantially different as FPPGAs do not carry digital logic operations but rather, they exploit optical interference to perform very high-speed analog operations acting over the phase and amplitudes of optical signals in a controlled environment provided by the chip's reduced footprint.

The Field Programmable Photonic Gate Array (FPPGA) [24] is an integrated photonic device/subsystem that operates in a similar way to a Field Programmable Gate array in electronics. Figure 1 provides an overall description of its main building blocks.

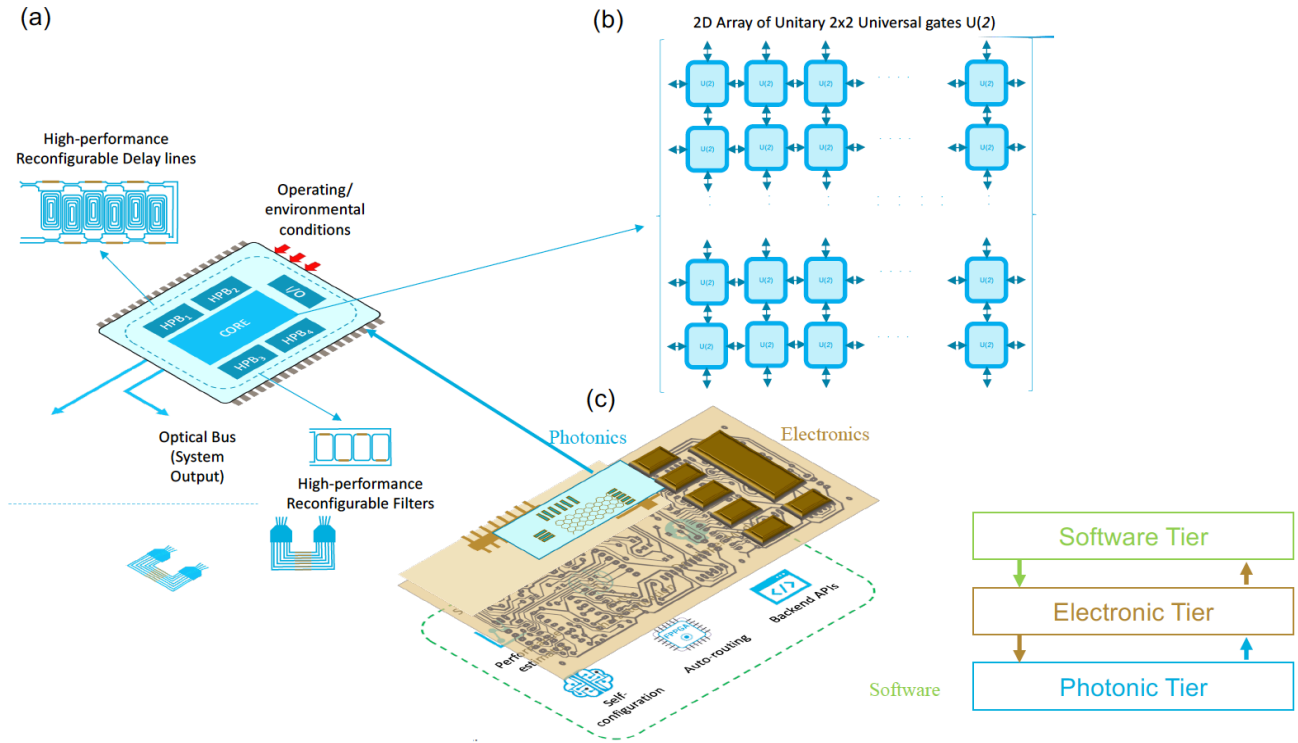

Figure 1. High-level layout of the FPPGA device. (a) Detail of the photonic tier, (b) Detail of the structure of the optical core formed by an array of interconnected programmable 2x2 unitary gates, (c) Overall device including the electronic and software tiers and their interdependencies.

Figure 1(a) shows a basic high-level architecture layout of the photonic Tier. It consists of a core composed of a 2D array of programmable universal 2x2 unitary gates U(*2*) implemented over a photonic chip and shown in Figure 1(b) and several external devices, which act either as input/output (I/O) ports or High-Performance Blocks (HPBs). In the core, the U(*2*) gates provide the building blocks for implementing basic optical analog operations (reconfigurable and independent power splitting and phase shifting). The full device, shown in Figure 1(c) incorporates as well an electronic and RF Tier in charge of driving and monitoring the photonic hardware as well as a software tier in charge of programming, optimizing, controlling and healing the device operation against undesired drifts and failures. In this paper we describe or recent work, advances and main considerations that have to be taken into account in the design of the different tiers involved in the successful operation of the device.

## 2. FPPGA PHOTONICS TIER

The core of a digital electronic FPGA hardware is built upon a set of interconnected configurable logic elements (CLEs), input/output devices and high performance blocks (HPBs) as shown in Figure 2(a). It is customary to employ elementary *irreversible* gates [25], [26] for bit processing using boolean logic as building blocks for CLEs [2] as shown. These gates are characterized by the fact that the number of input ports (or FAN-IN) is 2 while the number of output ports of FAN-

OUT is 1. Figures 2(b) and (c) show some basic examples of these gates, which are characterized by their so-called truth tables. The term irreversible reflects the fact that the input cannot be deduced from the output unambiguously. By cascading thousands of these gates one can build extremely complex combinatorial and sequential boolean circuits.

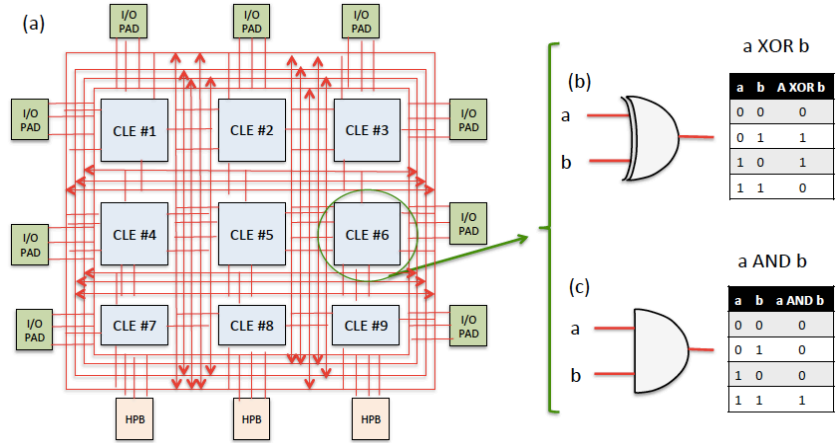

Figure 2. (a) Schematic layout of an electronic FPGA containing a set of interconnected Configurable Logic Elements (CLEs), Input/output Pads and High-performance Blocks (HPBs) and Examples of irreversible logic Boolean gates employed in the design of CLEs (b) XOR gate, (c) AND gate.

The core of a FPPGA on the contrary as shown in figure 1(b) is based on reversible and programmable [26], [27] unitary 2x2 gates U(*2*), which feature the same number of input and output ports and are characterized as well by truth tables. In this case, the input state can be deduced from the gate output as the gate operation can also be described by a unitary matrix transformation *U*. If ***I*** and ***O*** denote respectively the input and output vectors, the ***O=UI***, hence ***I=U⁻¹O***, but since ***U*** is unitary its inverse is given by the Hermitian conjugate. Reversible gates can be employed to perform digital Boolean operations but this process is inefficient compared to the use of irreversible gates as it entails the use of fixed or *ancilla* bits and produces as well garbage bits, which are not useful to the rest of the computation [28]. As a consequence, reversible gates are not employed in digital electronics. Reversible gates have found an application niche in the field of photonics quantum computation [13], [25-28]. The two main reasons are firstly that quantum computation does not rely on in Boolean logic but rather on the use of linear unitary transformations as a qubit $|\psi\rangle = \alpha|0\rangle + \beta|1\rangle$ resembles more an analog than a digital signal as the multiplying complex coefficients α and β states can be continuously changed. The second reason is that reversible gates can be engineered to exploit the garbage bits as heralding ports [26] to certify the correct operation of the gate. Since programmable integrated photonic circuits for classical signal processing applications handle analog signals as well, it makes sense therefore to consider the use of reversible gates as a basic building block to implement complex circuit structures just much in the same way as they are employed for quantum photonics.

To understand how arbitrary and programmable U(*2*) gates can be implemented in integrated photonics we start by considering the basic 2x2 Pauli matrices given by [26],[27]:

$$\sigma_o = I = \begin{pmatrix} 1 & 0 \\ 0 & 1 \end{pmatrix}; \quad \sigma_1 = X = \begin{pmatrix} 0 & 1 \\ 1 & 0 \end{pmatrix}$$
$$\sigma_2 = Y = \begin{pmatrix} 0 & -j \\ j & 0 \end{pmatrix}; \quad \sigma_3 = Z = \begin{pmatrix} 1 & 0 \\ 0 & -1 \end{pmatrix} \quad (1)$$

Two-dimensional rotation matrices (by an angle θ) around axes x, y and z are defined by the following transformations:

$$R_x(\theta)=e^{-j\frac{\theta}{2}X};\ R_y(\theta)=e^{-j\frac{\theta}{2}Y};\ R_z(\theta)=e^{-j\frac{\theta}{2}Z} \qquad (2)$$

Using basic operator theory, it can be readily shown that:

$$R_x(\theta)=\begin{pmatrix}\cos(\theta/2) & j\sin(\theta/2) \\ j\sin(\theta/2) & \cos(\theta/2)\end{pmatrix}$$

$$R_y(\theta)=\begin{pmatrix}\cos(\theta/2) & -\sin(\theta/2) \\ \sin(\theta/2) & \cos(\theta/2)\end{pmatrix} \qquad (3)$$

$$R_z(\theta)=\begin{pmatrix}e^{-j(\theta/2)} & 0 \\ 0 & e^{j(\theta/2)}\end{pmatrix}$$

These rotations can be implemented in the photonics domain respectively using a directional coupler, a 3-dB Mach-Zehnder tunable coupler (3dB-MZI) and two parallel waveguides including two phase shifters configured in push-pull mode respectively as shown in figure 3(a)-(c) where expressions for $\beta_1$, $\beta_2$, $\varphi_o$, $\varphi(\Delta\beta)$ can be found in [29]. Furthermore, rotation operators and their implementations are fundamental since by virtue of the Euler theorem [26], [27] any arbitrary 2x2 unitary matrix U(2) can be expressed in either of the two following sequential rotation matrix concatenation forms:

$$U=e^{j\delta}R_z(\alpha)R_y(\beta)R_z(\gamma)$$
$$U=e^{j\delta}R_z(\alpha)R_x(\beta)R_z(\gamma) \qquad (4)$$

Where $\alpha, \beta, \gamma$ and $\delta$ are a set of four real and independent numbers. This means that compact configurations consisting of a preceding and succeeding differential phase shifting units enclosing either a Dual-Drive Directional Coupler (DD-DC) [30] or a 3-dB-MZI unit as shown in figures 3(d) and 3(e) respectively can be employed to implement these arbitrary units. U(2) gate programmability is then achieved by means of electric signal biasing the different phase shifters.

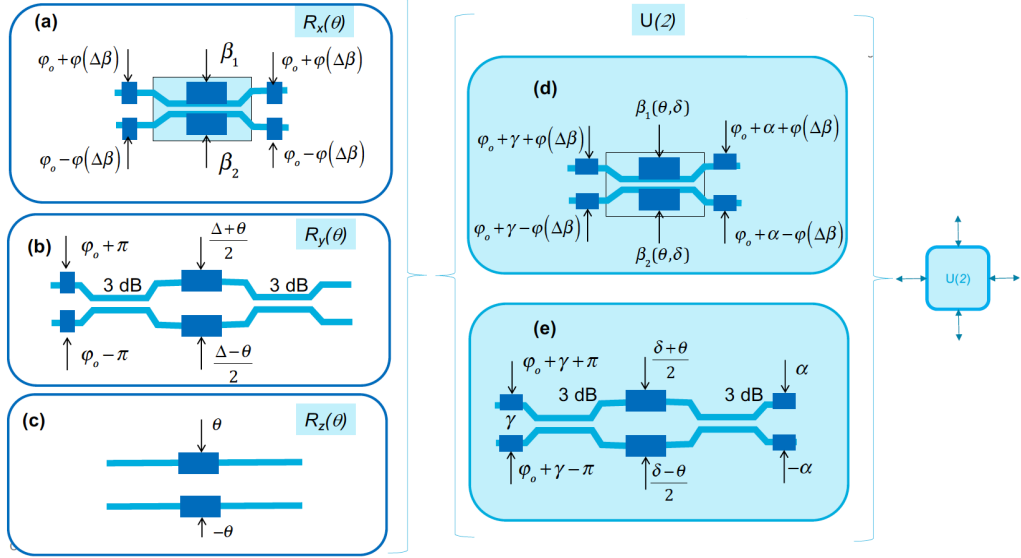

Figure 3. (a) Schematic layout for the photonic implementation of $R_x(\theta)$ based on a tunable directional coupler. (b) Schematic layout for the photonic implementation of $R_y(\theta)$ based on a tunable 3dB MZI coupler. (c) Schematic layout for the photonic implementation of $R_z(\theta)$ based on a differential push-pull phase shifter. (d) Integrated photonic implementation of an arbitrary unitary 2x2 matrix transform using a cascade of $R_x(\theta)$ and $R_z(\theta)$ rotations. (e) Integrated photonic implementations of an arbitrary unitary 2x2 matrix U(2) using cascades of $R_x(\theta)$ $R_y(\theta)$ and $R_z(\theta)$ rotations.

The 2D concatenation of these elements can follow different geometries, some of which are shown in figure 4(a). these can be readily implemented by square, triangular and hexagonal integrated waveguide meshes leveraging from the spatial isomorphisms shown in figure 4(b).

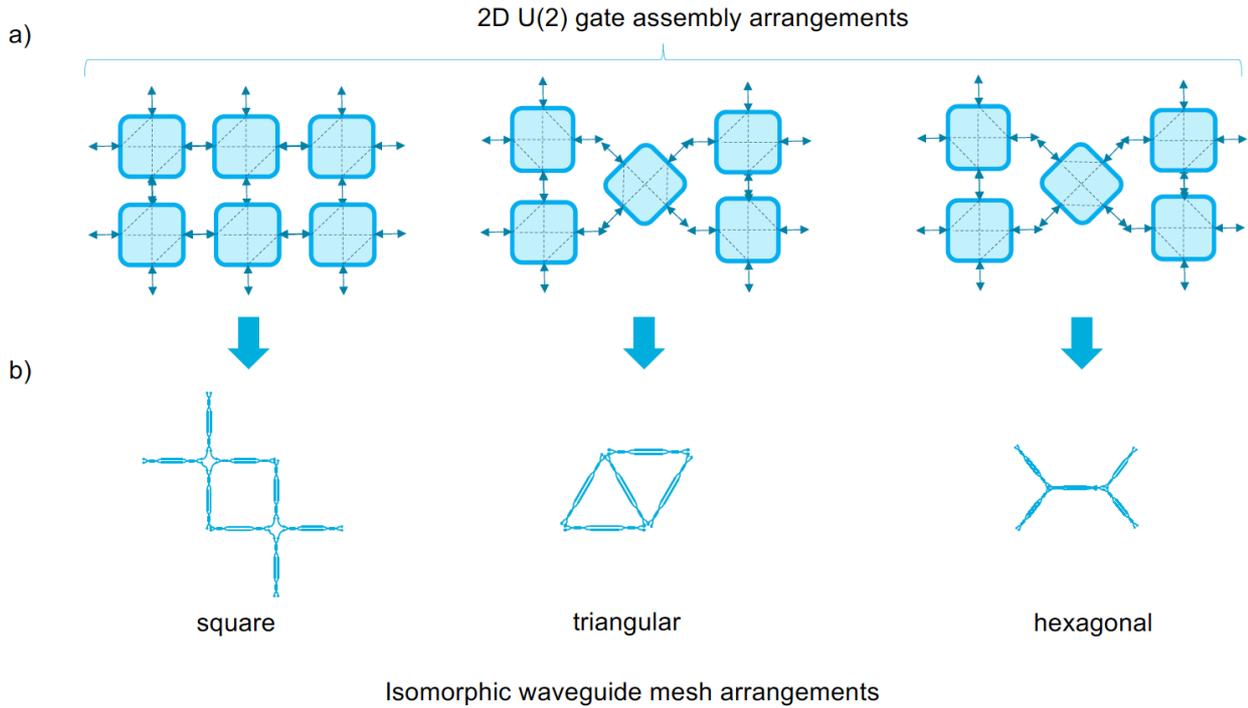

Figure 4. (a) 2D interconnection arrangements for U(2) gates. (b) spatial isomorphism between the interconnection arrangements and the square, triangular and hexagonal integrated waveguide mesh arrangements.

Regarding I/O ports, these can be implemented via grating or edge couplers, while HPB designs are specific to the targeted functionality envisaged for the block. As with electronic FPGAs, it is expected that the number, functionality and and interconnection schemes for HPBs will be specified by customers. FPPGA photonic hardware tier design faces several challenges, mostly connected to scalability and integration density. Some of these are briefly addressed in section 5.

## 3. FPPGA ELECTRONIC AND SOFTWARE TIERS

The programing and control of the FPPGA is provided by control electronics, and software framework. We briefly describe the most relevant issues related to them.

### 3.1 Electronic Tier

The electronic tier includes the hardware required to perform the monitoring from the on-chip optical readouts, to run the configuration algorithms and to drive the electro-optical actuators. Together with the PIC, they close a closed-loop control system, [31]. The subsystems employed are described as follows:

*Processing Unit:* The PU configures and sets the signals sent to the driving system circuitry. The configuration is chosen based on the desired application to be performed using the photonic circuit and the values provided by the software layer.

*Multi-channel Electronic Driving Array (MEDA):* The MEDA is the electronic subsystem that receives a digital signal from the PU. The said signal contains the configuration of every actuator in the system and translates it into the required driving signals. The format of these signals varies depending on the type of integrated actuator in the PIC. Typically, reconfigurable multi-channel current or voltage sources are employed for this purpose. The precision, linearity, stability, resolution and low-noise features are essential to ensure the correct performance of the PIC. The format of the signal varies from constant direct current driving [32] to digital pulsed formats [33].

*Multi-channel Electronic Monitoring Array (MEMA):* This subsystem obtains electronic readouts from the programmable PIC. Most of them carry signals that are proportional to the optical power (i.e. signals from opto-electronic transducers) in order to monitor the PIC behavior at specific points in the circuit. Others monitor different parameters like the thermistor in a temperature-controlling closed loop. The signal is then converted into a digitized format as required by the PU after which they are aggregated into an information array.

With regard to the optical power monitors, their number and location vary for each design. On-chip monitors enable a compact and small form-factor system. These can be classified in two approaches. First, the most conventional is the use of photo-diodes. Typically, such devices exhibit a responsivity in excess of 0.7 A/W along with dark currents less than 50 nA especially when used as low-speed, monitor elements. These can be placed either at the optical termination of the chip or elsewhere in the circuit preceded by an optical tapping element that extracts a portion of the signal of a waveguide and delivers it to the photo-diode. Depending on the power budget and the specifications, this might be a viable option. An alternative approach is the use of contactless integrated photonic probes (CLIPPs) [34], that provide a non-invasive solution with negligible loss.

### 3.2 Software Tier

The software tier can be classified in two additional levels. A lower level includes the basic routines to drive the actuators and to extract the readouts from the on-chip readouts. These basic routines are employed as an interface by more advanced routines that enable the programming of photonic integrated circuits. Among these routines, we can classify them depending on the prior information that require to work.

*Programming algorithms that require prior-knowledge*: These routines can provide programming of photonic circuits if they receive information about the status of the PIC. For example, this information can be the estimated loss associated to each component in the circuit, the electro-optical mapping of each integrated component, the power consumption required by each phase actuator and the interconnection mapping of the device. Examples of self-characterization routines providing this information has been proposed relying on periodic and iterative routines, [31]. With this information, auto-routing algorithms can be employed to optimize the hardware resources and to find the optimum circuits between optical nodes in the waveguide mesh arrangement. These can be optimized to perform a certain interconnection or delay line and to reduce the accumulated loss and power consumption [35, 36].

The auto-routing algorithm can be combined to configure presets of different photonic integrated circuits. This presents can be called by the *Processing Unit* to dynamically configure the FPPGA [37].

*Programming algorithms that do not require prior-knowledge:* Even employing self-characterization routines, gathering all the information of the photonic circuit becomes challenging for large scale circuits. For example, the tuning crosstalk (mainly thermo-optic crosstalk) in a dense photonic integrated circuit is impractical. In addition, some of the conditions might be susceptible to change after the characterization. To solve this, a set of routines employ optimization methods where an application-specific cost function is minimized through the variation of the phase actuators, assuming the circuit as a black box. Thus, the optimization procedure considers all the non-ideal and environmental effects, providing solutions that mitigate them [38]

## 4. FPPGA APPLICATIONS

Programmable photonics can find applications in a myriad of areas, as we will illustrate in this section. Interestingly, one of the first fields in which it has had an earlier penetration is that of Quantum Information that encompasses, communications, computing, sensing and tomography. Proof of many of these topics using lightwave technology require the use of reconfigurable linear optics transformations and at earlier stages these have been implemented resorting to large scale bulk optics setups, which prevent the development of more complex and scalable quantum optics configurations requiring tens or hundreds of modes. On top of this, linear optic systems with the required fidelity require a strict control of interference though demanding phase stability mechanisms. Integrating a considerable number of photonic elements on a chip in order to implement multiport interferometers as described in sections 4 and 6 has been identified as the only viable path leading towards quantum information systems at a technological reach. Programmability brings the added value of enabling successive implementations of quantum optics experiments featuring random variation in the parameters of interest. We provide a brief introduction to the applications of programmable photonics to quantum information systems. Though the applications are many we focus in particular our attention to linear optical quantum gates and quantum transport simulation. Other applications such as boson sampling and complex Hadamard transformations the reader is directed to the literature [12], [39],[40].

The second and probably wider area of application of programmable photonics is in classical systems. While so far most of these systems have relied mainly on pure fixed application specific photonic circuits, the advantages brought by programmability can become decisive when these systems grow in complexity. Some of these application areas where programmable photonics can become the pervasive enabling technology in the near future are reviewed in [41]. For example, in optical switching interconnection and routing, programmable photonics provides a myriad of powerful solutions that encompass not only pure spatial switching and routing, but also broadcasting and wavelength selective operation. A second emerging field of application is Artificial Intelligence and neurocomputing. Programmable photonics enables the implementation of reconfigurable analog cores required for the operation of neural networks and neurophotonic systems. Recent works have reported first proof of concept results showing very promising outputs. Microwave and analog photonics is a particularly appealing area for programmable photonics as it can provide substantial cost reductions and broadband operation flexibility by enabling the possibility of implementing most of the required functionalities in a single chip. The use of these functionalities spans a huge area of applications including 5G communications, satellite payloads, electromagnetic and photonic radar, Internet of Things, and autonomous driving, among others.

Figure 5 (a) illustrates a possible application of FPPAs to reconfigurable coherent transceivers. The available real estate of the device is sectioned into two different parts. The upper part implements the TRX stage where a HPB providing the continuous-wave ligtwave carrier is first filtered by one or two (in Vernier configuration) ring cavities to provide extra reduction in linewidth and phase noise (note that the inclusion of one or the two ring cavities can be programmed as desired). The filtered CW signal is then sent to an external HPB implementing the I-Q modulation before exiting the TRX. In the receiving section the input signal is combined with a local oscillator (LO) provided by another HPB in a reconfigurable 2x4 hybrid that is implemented by the waveguide mesh. Here the values of the hybrid 2x4 transfer matrix can be corrected if needed to provide a complete balanced signal (a feature which is not possible in current receivers). Note that the linewidth of the LO can be improved by implementing a ring cavity filter before its input to the 2x4 hybrid, although this option is not illustrated in figure 5. The internal programming of the hexagonal waveguide mesh implementing the FPPGA core is shown in figure 5(b). Further improvements and additional features can be of course implemented to the transceiver.

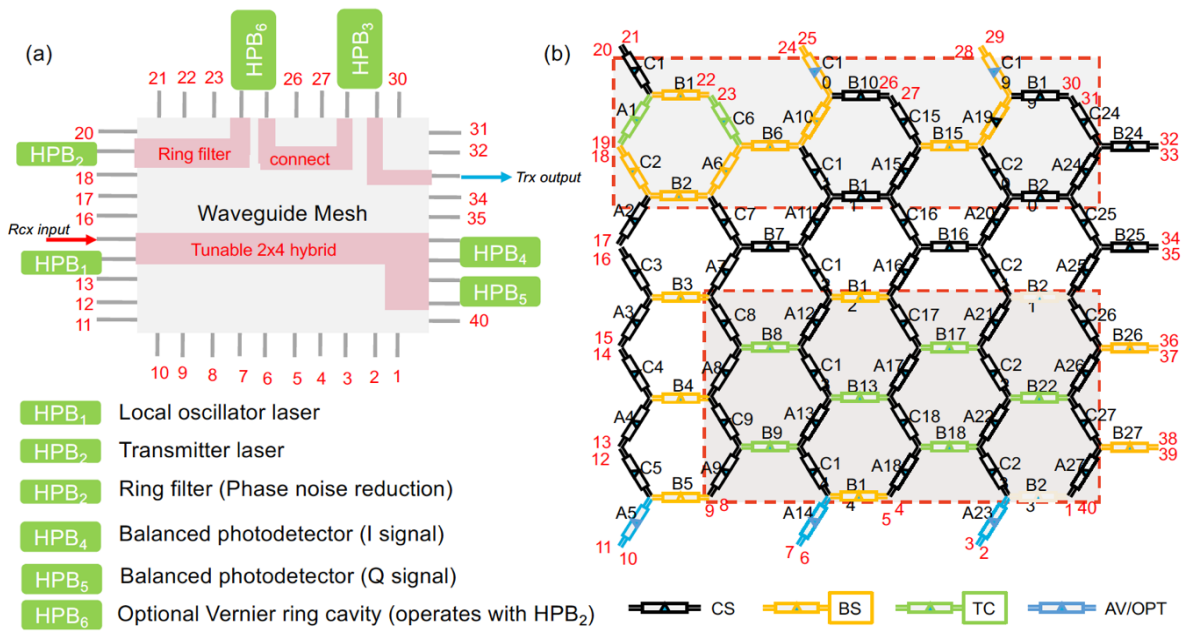

Figure 5. (a) An example of the implementation of a reconfigurable coherent transceiver by means of a FPPGA device. (b) internal programming of the FPPGA hexagonal waveguide mesh core.

## 5. FPPGA EVOLUTION ROADMAP

Current FPPGA demonstrations integrate a few tens of phase actuators and their software capabilities lied have not been demonstrated employing real large-scale hardware. Although the FPPGAs are expected to achieve the integration of 1000 phase actuators in the first five years, key enabling technology will be required in order to achieve this goal, efficiently. First, developing power-efficient mechanisms for the implementation of phase-shifters, possibly using non-volatile approaches will be required to maintain a low power consumption and to enable the integration of the driving system [42]. In addition, developing efficient schemes for the monitoring and control hardware and software will be essential to supervise and manage an extensive number of TBUs in real time and improve the reconfiguration times. Moreover, another key challenge for the hardware scalability is the electrical interfacing of a large number of signals. Solutions might come from the monolithic co-integration of photonics and electronics [43] or by employing flip-chip connections to a PCB. Although different in application, recent published chips with similar requirements integrate a great number of optical power monitors and phase shifters. For example, two recent demonstrations integrate 900 optical power monitors and 448 phase shifters [44], and 1024 phase shifters [45], respectively in footprints lower than 300 mm$^2$.

Most of these challenges can be addressed by the software layer. Thus, it is expected that the use of self-configuration routines can manage efficiently the use of imperfect component while providing fault-tolerant programming of the FPPGA. Smarter routines are expected to optimize the resources of the FPPGA more efficiently, providing the simultaneous configuration of circuits at the same time [38].

Future evolution will include high-speed electronic components to perform arbitrary microwave photonic functionalities and smart transceiver operations.

## 6. SUMMARY AND CONCLUSIONS

We have reviewed some of the basic principles, fundamentals, technologies and architectures for the implementation of Field Programmable Photonic Field Arrays (FPPGAs). Programmable photonics is in a way following similar steps as programmable electronics did in the last two decades of the last century. Figure 6 shows a time frame evolution of both fields, where some salient milestones are outlined.

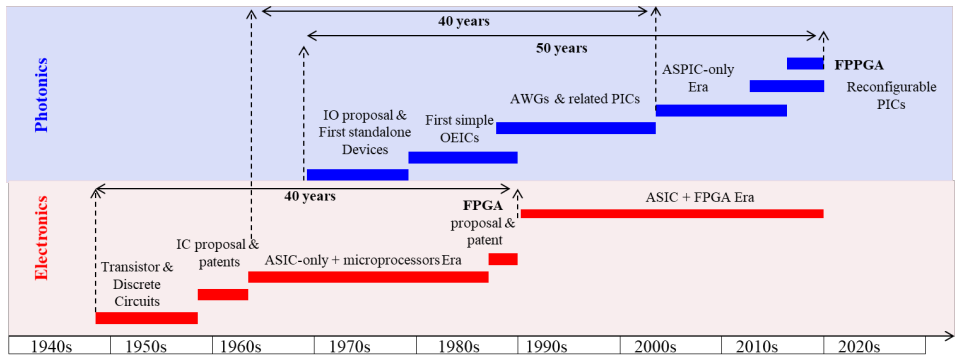

Figure 6. Significant milestones and evolution periods in integrated electronics and photonics

While the similarity of the main evolution paths can be identified, it is important to acknowledge that there are as well important differences. The first important difference is that signal processing in PIP is analog as opposed to the digital character of integrated electronic chips. The lack of optical memories precludes digital operations for which a previous analog to digital conversion plus bit storage is required. All optical signal processing has to be performed on the fly, as the analog signal propagates. Rather than being a limitation, this should be considered as a complementary feature brought by photonics which, if conveniently leveraged and complemented to the digital nature of electronics can open a completely new signal processing paradigm where both worlds cooperate synergistically to benefit a given application field. One example is artificial intelligence, where computing paradigms based on analog signal processing are being now considered as a means to emulate basic operations carried out by neurons, which are complementary and outperform some of those carried by digital processing.

To fully exploit the capabilities of FPPGA technology further research is needed in different areas:

a) The development of a theoretical background for analog optical gates based on a SU(*2*) algebra of rotation matrices that can play a similar role as Boole algebra in digital electronics.
b) Further investigation on efficient decompositions of complex multiport interferometric and waveguide mesh architectures in terms of SU(*2*) gates.
c) Developing arbitrary circuit synthesis algorithms for FPPGA structures.
d) Developing power-efficient mechanisms for the implementation of phase-shifters, possibly using non-volatile approaches.
e) Developing monitoring and control hardware and software capable of supervising an extensive number of TBUs in real time.
f) Developing routing and placement software to optimize the allocation of the programmed circuit/s within the available real estate in the chip and reroute connections in case of one or multiple TBU failures.
g) Investigation on the alternatives to scale interferometers and waveguide meshes by increasing the number of TBUs. In particular by adding active elements for loss compensation, by the reduction of the TBU footprint and by reducing the insertion losses of the passive elements.
h) Developing the technology to co-integrate FPPGAs with sophisticated programming and control electronics.